\documentclass[%
 aip,
 chaos,%
 amsmath,amssymb,
 reprint,%
]{revtex4-1}

\usepackage{graphicx}

\usepackage{color}

\begin{document}

\title{
Josephson phase diffusion  in the SQUID ratchet}

\author{Jakub Spiechowicz}
\affiliation{Institute of Physics, University of Silesia, 40-007 Katowice, Poland}

\author{Jerzy {\L}uczka}
\email{jerzy.luczka@us.edu.pl}
\affiliation{Institute of Physics, University of Silesia, 40-007 Katowice, Poland}
\affiliation{Silesian Center for Education and Interdisciplinary Research, University of Silesia, 41-500 Chorz{\'o}w, Poland}


\begin{abstract}
We study diffusion of the Josephson phase in the asymmetric SQUID subjected to a time-periodic current and pierced by an external magnetic flux. We analyze a relation  between phase diffusion and quality of transport characterized by the dc voltage across the SQUID and efficiency of the device.   In doing so, we concentrate on the previously reported regime [J. Spiechowicz and J. {\L}uczka, New J. Phys. \textbf{17}, 023054 (2015)] for which efficiency of the SQUID attains a global maximum. For long times, the mean-square displacement of the phase is a linear function of time, meaning that  diffusion is normal. Its coefficient is small indicating rather regular  phase evolution. However, it can be magnified \emph{several times} by tailoring experimentally accessible parameters like amplitudes of the ac current or external magnetic flux. Finally, we prove that in the deterministic limit this regime is essentially \emph{non-chaotic} and possesses an unexpected simplicity of attractors.
\end{abstract}

\pacs{
74.25.F-, 
85.25.Dq, 
05.40.-a, 
05.60.-k. 
}

\maketitle

\begin{quotation}
A Superconducting quantum interference device (SQUID) is one of the most important element of apparatuses  of research laboratories worldwide. It exhibits a wide variety of phenomena and has been successfully used for testing the fundamentals of quantum mechanics, quantum information and chaotic phenomena in classical physics. In the semiclassical regime, dynamics of the Josephson phase in the SQUID can be visualized as a Brownian particle moving in a periodic potential. This correspondence has allowed to study transport properties of the SQUID by applying methods of Langevin equation. Earlier research in this field has been concentrated on the influence of thermal noise and external driving on   current-voltage characteristics. The present paper studies diffusion of the Josephson phase. We consider a special type of the SQUID which operates as a Brownian ratchet. In particular, we analyze the phase diffusion in the regime for which the efficiency of the SQUID is globally maximal. 
\end{quotation}

\section{Introduction}
\label{intro}

The SQUIDs \cite{squid1} are elements of ultrasensitive electric and magnetic measurement systems. They are also exploited in a number of commercial applications in industrial metrology, geophysical systems and medicine as a noninvasive technique for investigating  the  human body. \cite{fagaly2006, matti1993} In terms of the Stewart-McCumber model,\cite{stewart,mccumber} both  the dynamics of the Josephson phase in the SQUID and the position of a Brownian particle can be described by similar Langevin equation. \cite{kautz1996} 
 Therefore time evolution of the Josephson phase is analogous to a random motion of a Brownian particle in a spatially periodic potential. While prior studies have mostly been focused on the current-voltage characteristics, diffusion of the Josephson phase, i.e. its mean-square displacement, has not been intensively analyzed. There are only several papers closely or loosely 
related to this latter subject.\cite{gang1996, blackburn1996, harish2002, tanimoto2002, guo2014} 

In this work, we study diffusion of the Josephson phase in the  asymmetric SQUID \cite{zapata1996prl, weiss2000, sterck2002, berger2004, sterck2005, sergey, sterck2009} which is composed of three capacitively and resistively shunted Josephson junctions. Two junctions are collocated in series in one half-piece of the ring and the third junction is disposed in the other half of the ring. The SQUID is driven by a time-periodic current and subjected to an external constant magnetic flux. Its dynamics is extremely rich and complex even in the deterministic case, involving  harmonic, subharmonic, quasiperiodic and chaotic trajectories. At non-zero temperature, thermal fluctuations lead to diffusive behaviour with random escape events among possibly coexisting attractors   and with the system typically exploring its whole phase space. 
Due to a multi-dimensionality of the parameter space describing the model it would be extremely difficult to perform a complete analysis of diffusion process of the Josephson phase in such a setup. Therefore we consider it only in the previously reported regime \cite{NJP15} for which the Stokes efficiency of the SQUID attains its global maximum. 

The paper is structured as follows. In Sec. \ref{model} we present a model of the asymmetric SQUID in terms of the Langevin equation for the Josephson phase. Next, we define several quantifiers characterizing the quality of transport process occurring in this setup. Among them are those that relate directly to an asymptotic long time stationary average voltage drop like its variance and the Stokes efficiency of the SQUID. However, we also study in this context those connected with the realization of stochastic motion of the Josephson  phase like its mean squared displacement and a diffusion coefficient. We also combine these complementary quantities into the  so called P\'{e}clet number. In Sec. \ref{results} we present results based on the comprehensive numerical simulation of the studied system in a regime for which the transport efficiency is globally maximal. In particular, we demonstrate a possibility of steering of  diffusive behaviour of the phase by tuning the experimentally accessible parameters like the amplitude of ac current applied to the system or the external magnetic flux. It turns out that the diffusion coefficient can be significantly enhanced by small variation of these quantities. Moreover, we establish a clear relation between the directed transport, its efficiency and the diffusive motion of the Josephson  phase. It is very important that the regime of globally maximal transport quality measured by the Stokes efficiency is also effective in the sense that phase motion is ordered and regular. Finally, the last section provides summary and conclusions.
\begin{figure}[t]
    \centering
    \includegraphics[width=0.9\linewidth]{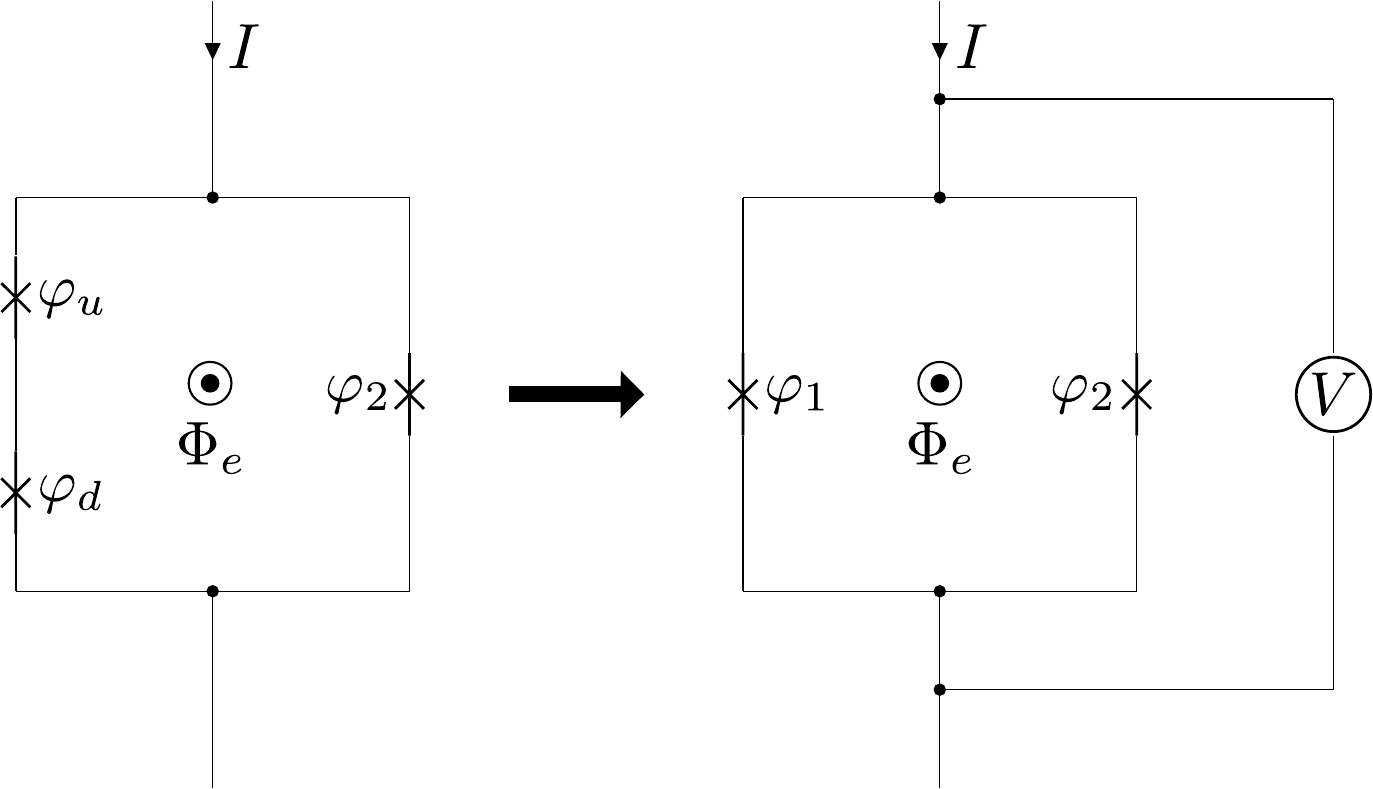}
 \caption{Schematic asymmetric SQUID composed of three Josephson junctions and the equivalent circuit built from two junctions. The Josephson phase difference is $\varphi_1 = \varphi_u + \varphi_d$, the externally applied current is $I$, the external magnetic flux is $\Phi_e$ and the instantaneous voltage across the SQUID is \mbox{$V=V(t)$}.}
    \label{fig1}
\end{figure}

\section{Model of the asymmetric SQUID}
\label{model}
The asymmetric SQUID  \cite{zapata1996prl, weiss2000, sterck2002, berger2004, sterck2005, sergey, sterck2009} is presented in Fig. \ref{fig1}. It is formed by a superconducting loop with two resistively and capacitively shunted Josephson junctions \cite{josephson} in the left arm and only one in the right arm. The crosses denote the junctions and $\varphi_k \equiv \varphi_k(t)$ ($k=u, d, 1, 2$) are the phase differences across them. Each junction is characterized by the capacitance $C_k$, resistance $R_k$ and critical Josephson current $J_k$, respectively. To reduce a number of parameters of the model, we consider only a special case when two junctions in the left arm are identical, i.e. $J_u = J_d \equiv J_1, R_u = R_d \equiv R_1/2, C_u = C_d \equiv 2C_1$. In some regimes \cite{spiechowicz2014} they can be considered as one for which the supercurrent-phase relation takes the form $J_1 \sin{\left(\varphi_1/2\right)}$, where $\varphi_1 = \varphi_u + \varphi_d$. Additionally, the SQUID is threaded by an external magnetic flux $\Phi_e$. As a consequence the effective potential experienced by the  phase $\varphi_1$  forms a ratchet structure.  \cite{hanggi2009} The device is driven by an external current $I = I(t)$ which is composed of the static dc current $I_0$ and the ac component of amplitude $A$ and angular frequency $\Omega$, namely, 
\begin{equation}
\label{eq1}
    I(t) = I_0 + A \cos(\Omega t). 
\end{equation}
The Langevin equation for the phase $\varphi \equiv \varphi_1$ is of the form  \cite{spiechowicz2014}
\begin{equation}
    \label{eq2}
	\frac{\hbar}{2e} C \ddot{\varphi} + \frac{\hbar}{2e} \frac{1}{R} \dot{\varphi} + J(\varphi) = I(t) +	\sqrt{\frac{2k_B T}{R}}\,\xi(t),
\end{equation}
where the effective supercurrent $J(\varphi)$ reads
\begin{equation}
    \label{eq3}
    J(\varphi) = J_1\sin{\left( \frac{\varphi}{2} \right)} + J_2\sin{(\varphi + \tilde{\Phi}_e)}. 
\end{equation}
The parameters are: $C = C_1 + C_2$, $R^{-1} = R_1^{-1} + R_2^{-1}$, 
$k_B$ is the Boltzmann constant, $T$ is temperature of the system and ${\tilde{\Phi}_e}= 2\pi \Phi_e/\Phi_0$ is the dimensionless external magnetic flux. Thermal fluctuations are modelled by $\delta$-correlated Gaussian white noise $\xi(t)$ of the statistics:
\begin{equation}
    \label{eq4}
    \langle \xi(t) \rangle = 0, \quad \langle \xi(t)\xi(s) \rangle = \delta(t-s).
\end{equation}
\begin{figure}[t]
    \centering
    \includegraphics[width=0.9\linewidth]{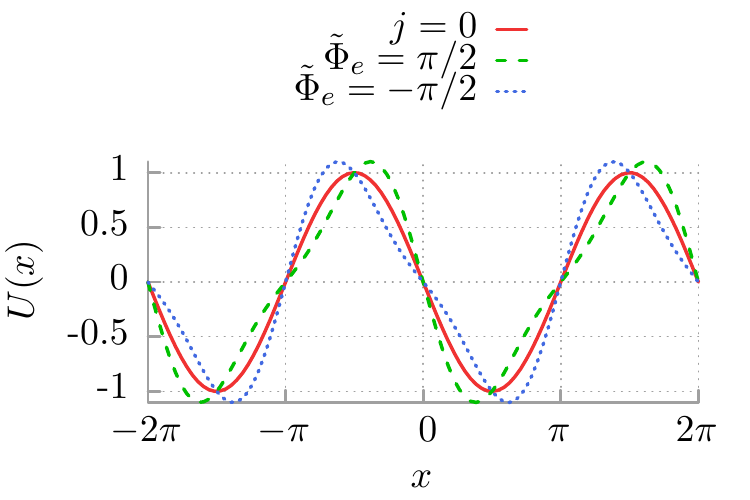}
    \caption{The potential (\ref{eq7}) for the symmetric case $j=0$ (solid red line) in comparison with the ratchet potential for $j=1/2$ and two values of the external magnetic flux $\tilde\Phi_e = \pi/2$ (dashed green line) and $\tilde\Phi_e = -\pi/2$ (dotted blue line).}
    \label{fig2}
\end{figure}
The average voltage across the device can be calculated from the relation 
\begin{equation}
\label{Vol}
\langle V \rangle = (\hbar/2e)\langle \dot{\varphi}_1 \rangle = (\hbar/2e) \langle \dot{\varphi}_2 \rangle,  
\end{equation}
where the averaging is over the period $2\pi/\Omega$ of the ac current. This relation is also valid 
in the long time regime when additionally averaging is performed over initial conditions and all realizations of thermal noise. 

It is useful to interpret Eq. (\ref{eq2}) in the mechanical framework as a model of the inertial Brownian particle subjected to the conservative force $J(\varphi)$ and propelled by the time dependent driving $I(t)$. In this correspondence, the particle position $x$ translates to the phase $\varphi$, its velocity $v= \dot{x}$ to the voltage $V$, the conservative force to the supercurrent $J(\varphi)$, the external force to the current $I(t)$, the mass $m$ to the capacitance $C$ and the friction coefficient $\gamma$ to the normal conductance $G=1/R$.

In our recent paper \cite{spiechowicz2014} we analyzed conditions that are necessary for  generation and control of the voltage drop across the device. In particular, we focused on the direction and magnitude of transport as well as its dependence on the system parameters. We found the intriguing features of a negative absolute and differential conductance,\cite{machura2007, speer2007, kostur2008} repeated voltage reversals \cite{kostur2000} and noise induced voltage reversals. \cite{kula1998} We showed how the direction of transport can be controlled by the applied magnetic field. Moreover, very recently \cite{NJP15} important aspects concerning the quality of transport occurring in this system have been addressed. We analyzed  fluctuations of the voltage and  energetics of the device. \cite{machura2004, machura2005, machura2006, kostur2006, machura2010} It turned out that the power delivered by the external current depends not only on  its amplitude and  frequency but also on thermal noise and the external magnetic field. \cite{jung2011} We explored a set of the system parameters to reveal a regime where the voltage rectification efficiency is globally maximal. We detected the surprising feature of the thermal noise enhanced efficiency \cite{spiechowicz2014pre} and showed how the efficiency of the device can be tuned by adjusting the external magnetic flux. However, apart from these well investigated problems there are still some questions about the transport properties of the system which should be imposed and answered. One important example might be a diffusion process \cite{gang1996, blackburn1996, harish2002, tanimoto2002, guo2014} of the Josephson phase. Therefore in this paper we focus on the connection between the directed transport quantified by the averaged voltage $\langle V \rangle$, its quality measured by the efficiency $\eta$  of the device and  diffusion of the Josephson phase  $\varphi$.  
\begin{figure*}[t]
    \centering
    \includegraphics[width=0.33\linewidth]{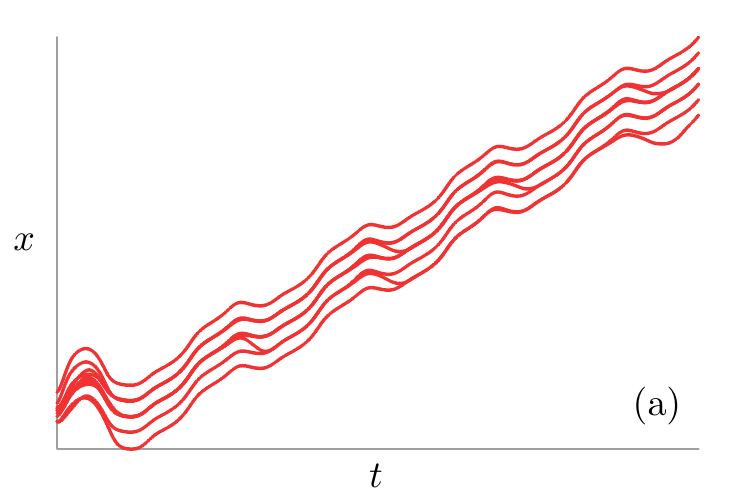}
    \includegraphics[width=0.33\linewidth]{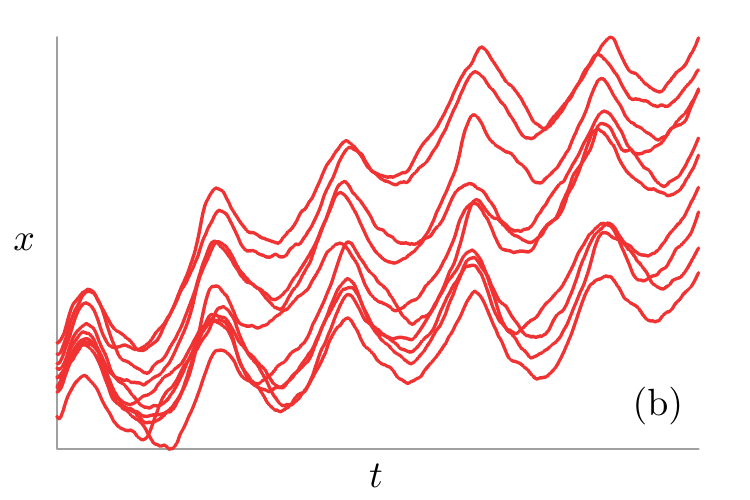}
    \caption{Two sets of illustrative trajectories of the phase motion $x(t)$ across the asymmetric SQUID device. Parameters are $\tilde{C} = 0.496$, $a = 1.55$, $\omega = 0.406$, $\tilde{\Phi}_e = \pi/2$, $j = 0.5$. In panel (a) and (b) the thermal noise intensity is set to $D = 10^{-5}$ and $D = 0.1$, respectively. A distinct different diffusive behaviour is observed.}
    \label{trajs}
\end{figure*}

\section{Transport quantifiers}
It is convenient to convert Eq. (\ref{eq2}) into its dimensionless form. There are several forms of such an equation depending on time scaling. Here, we follow Ref. \onlinecite{zapata1996prl} and define the new phase $x$ and the dimensionless time $\hat{t}$ as
\begin{equation}
	\label{eq5}
	x = \frac{\varphi + \pi}{2}, \quad \hat{t} = \frac{t}{\tau_c}, \quad \tau_c = \frac{\hbar}{eRJ_1}. 
\end{equation}
In these new variables, Eq. (\ref{eq2}) reads
\begin{equation}
    \label{eq6}
    \tilde C \ddot{x}(\hat{t}) + \dot{x}(\hat{t}) = -U'(x(\hat{t})) + F + a\cos(\omega \hat{t}) + \sqrt{2D}\,\hat{\xi}(\hat{t}),
\end{equation}
where the dot and prime denotes a differentiation with respect to the dimensionless time $\hat{t}$ and the phase $x$, respectively. The dimensionless capacitance $\tilde C$ is the ratio between two characteristic time scales $\tilde C = \tau_r/\tau_c$, where the relaxation time is $\tau_r = RC$. Other re-scaled parameters are $F = I_0/J_1$, $a = A/J_1$ and $\omega = \Omega\tau_c$. The periodic potential $U(x)$ of period $2\pi$ takes the form \cite{spiechowicz2014}
\begin{equation}
    \label{eq7}
    U(x) = - \sin(x) - \frac{j}{2} \sin(2x + \tilde\Phi_e - \pi/2), 
\end{equation}
where $j=J_2/J_1$. This potential is symmetric if there exists $x_0$ such that $U(x_0 + x)= U(x_0 - x)$ for any $x$. If $j \neq 0$, it is generally asymmetric and is called a ratchet type, see Fig. \ref{fig2}. However, even for $j \neq 0$ there are certain values of the external flux $\tilde \Phi_e$ for which it is still symmetric.  The rescaled zero-mean Gaussian white noise $\hat{\xi}(\hat{t})$ has the auto-correlation function \mbox{$\langle \hat{\xi}(\hat{t})\hat{\xi}(\hat{s}) \rangle = \delta(\hat{t}-\hat{s})$} and its intensity $D = e k_B T/\hbar J_1$ is the quotient of the thermal and the Josephson coupling energy. From now on we will use only the dimensionless quantities and therefore we skip all hats appearing in (\ref{eq6}).

There are several quantifiers characterizing transport properties of the system. The  most important are the current-voltage curves in the asymptotic limit of long times when all effects due to initial conditions and transient processes have quiet down. They can be obtained from Eq. (\ref{Vol}), the dimensionless form of which reads \cite{jung1993}
\begin{equation}
	\label{eq8}
	\langle v \rangle = \lim_{t\to\infty} \frac{\omega}{2\pi} \int_{t}^{t+2\pi/\omega} \mathbb{E}[\dot{x}(s)] \, ds,
\end{equation}
where $\mathbb{E}[\dot{x}(s)]$ denotes  averaging  over initial conditions and all realizations of  thermal noise.
 The stationary dimensional  voltage is then given as
\begin{equation}
    \label{eq9}
    \langle V \rangle  = R J_1 \langle v \rangle.
\end{equation}
The  long time average voltage $\langle v \rangle$ represents the basic transport measure. However, there are other transport quantifiers  like the \emph{voltage variance}
\begin{equation}
    \label{eq10}
    \sigma^2_v = \langle v^2 \rangle - \langle v \rangle^2, 
\end{equation}
which describes voltage fluctuations  around its average value $\langle v \rangle$.  The voltage drop $v(t)$ across the SQUID typically ranges within the interval of standard deviations,   
\begin{equation}
    \label{eq11}
    v(t) \in [ \langle v \rangle - \sigma_v, \langle v \rangle + \sigma_v ].
\end{equation}
It means that when $\sigma_v > |\langle v \rangle|$ the instantaneous voltage $v(t)$ may assume the opposite sign to the average voltage $\langle v \rangle$ and transport in not effective. 

Next, we can introduce  a measure for the efficiency of the SQUID in terms of the rectification of thermal fluctuations. It is defined as the ratio between the energetic output  and the input power $\langle [F+A\cos(\omega t)] v(t)\rangle$.   This quantity follows from an energy balance of the underlying Langevin equation  (\ref{eq6}) and  corresponds to the well known \emph{Stokes efficiency} \cite{wang2002, machura2004, wang2009}
\begin{equation}
    \label{eq12}
    \eta = \frac{\langle v \rangle^2}{\langle v \rangle^2 + \sigma_v^2 - D/\tilde{C}} = \frac{\langle v \rangle^2}{\langle v^2 \rangle - D/\tilde{C}}.
\end{equation}
One should note that this definition yields a non vanishing rectification efficiency even when there is no external static current, $F=0$. Furthermore, it follows that this result is accordant with our intuition: a decrease of the voltage variance $\sigma_v^2$ leads to an increase of the rectification efficiency. Consequently, to optimize the effectiveness of the device one should seek for regimes that \emph{maximize} the directed voltage and \emph{minimize} its fluctuations.

The next quantifier  characterizing transport is  related to   spread of trajectories of the stochastic phase  $x(t)$.  It is the  mean square displacement  \cite{atbook} of the phase defined as
\begin{equation}
    \label{eq13}
    \langle \Delta x^2(t) \rangle = \left\langle [ x(t) - \langle x(t) \rangle ]^2 \right\rangle  \nonumber  = \langle x^2(t) \rangle - \langle x(t) \rangle^2.
\end{equation}
%
One can expect that  in the long time regime it grows  according to a power law  \cite{atbook}
\begin{equation}
    \label{eq14}
    \langle \Delta x^2(t) \rangle_{\xi} \,\sim 2 D_xt^{\alpha}, 
\end{equation}
where the constant prefactor $D_x$ is sometimes called a diffusion coefficient and the exponent $\alpha$ characterizes a type of diffusion: \cite{atbook, metzler2000, bouchaud1990}  subdiffusion for $0 < \alpha < 1$, normal difussion for $\alpha =1$ and superdiffusion for $\alpha > 1$. Another special case is ballistic diffusion for $\alpha = 2$. When  $\alpha =1$ the diffusion coefficient can be  determined as
\begin{equation}
    \label{eq15}
    D_x = \lim_{t \to \infty} D_x(t) = \lim_{t \to \infty} \frac{\langle \Delta x^2(t) \rangle}{2t}.
\end{equation}
Otherwise the above definition is not constructive because such a  quantity is either zero (subdiffusion) or diverges to infinity (superdiffusion). The diffusion coefficient measures the spreading of trajectories $x(t)$ around its mean value $\langle x(t)\rangle$, see Fig. \ref{trajs}. Intuitively, when it is small  then the spread of trajectories is small and transport is more optimal.
\begin{figure*}[t]
    \centering
    \includegraphics[width=0.33\linewidth]{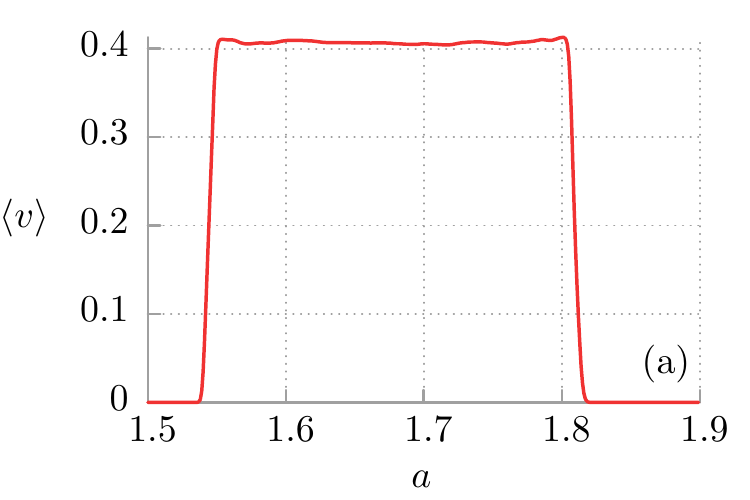}
    \includegraphics[width=0.33\linewidth]{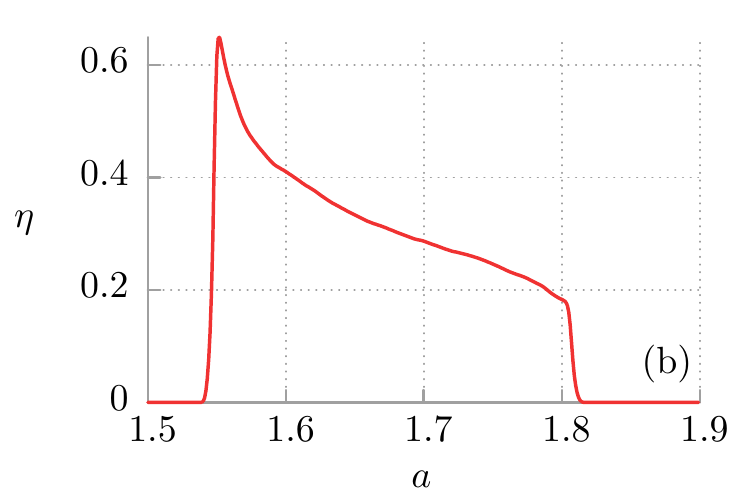} \\
    \includegraphics[width=0.33\linewidth]{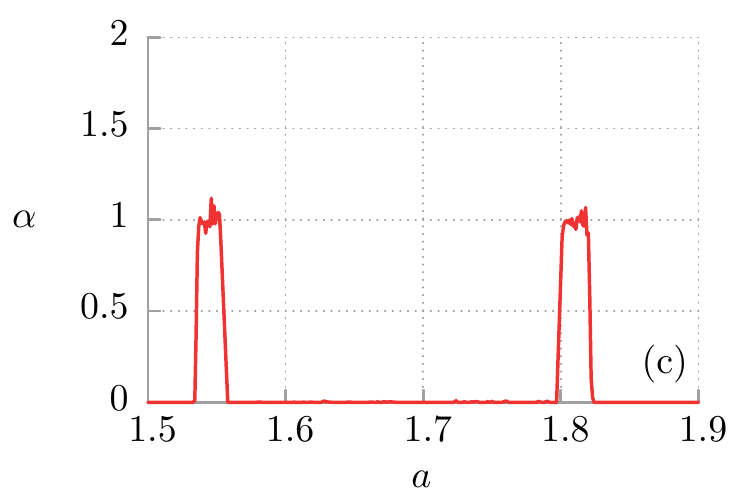}
    \includegraphics[width=0.33\linewidth]{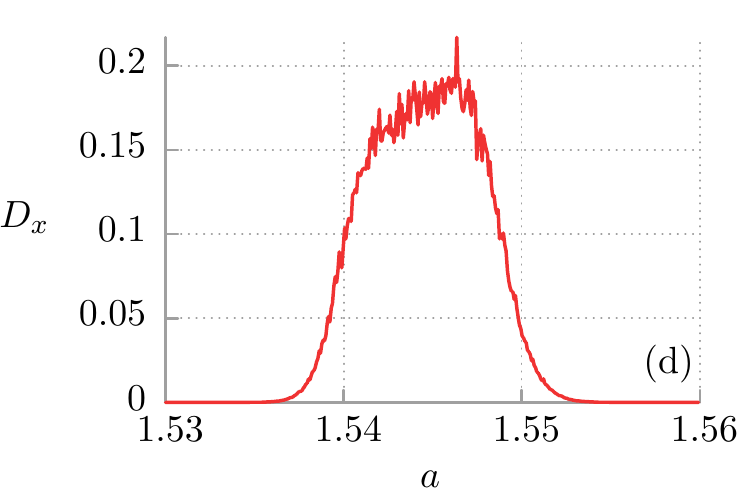}
    \caption{Impact of  the ac current amplitude $a$ on transport  characteristics.  (a): the dc voltage drop $\langle v \rangle$ across the SQUID, (b): the Stokes efficiency $\eta$, (c): phenomenological power exponent $\alpha$ describing a type of phase diffusion  and (d): the diffusion coefficient $D_x$ in the regime when $\alpha \approx 1$. Other parameters correspond to the regime for which the rectification efficiency $\eta$ takes its globally maximal value, namely $\tilde{C} = 0.496$, $a = 1.55$, $\omega = 0.406$, $D = 10^{-5}$, $\tilde{\Phi}_e = \pi/2$, $j = 0.5$ and  $F=0$.}
    \label{fig3}
\end{figure*}
\begin{figure}[b]
    \centering
    \includegraphics[width=0.7\linewidth]{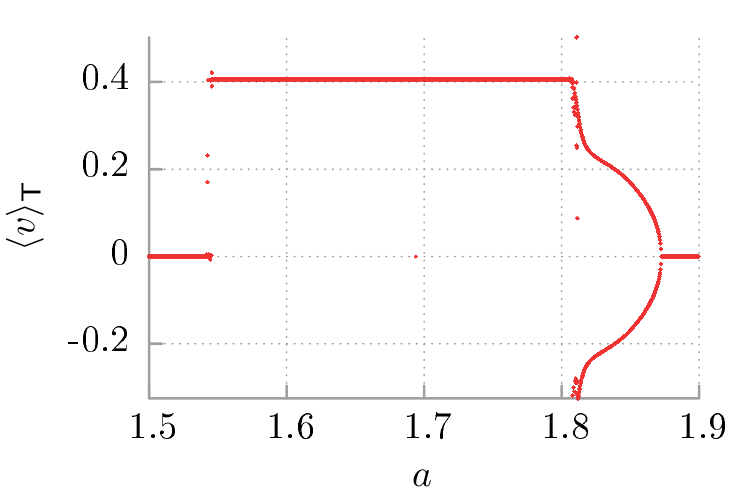}
    \caption{The dc voltage bifurcation diagram in the deterministic limit $D = 0$ as a function of the ac driving amplitude $a$. The remaining parameters are given in Fig. \ref{fig3}.}
    \label{fig4}
\end{figure}

Another way of introducing the diffusion coefficient is based on the generalized Green-Kubo  response theory.\cite{kubo1966, machura2005, machura2006} The asymptotic long time voltage autocorrelation function
\begin{equation}
    \label{eq16}
    C(t, \tau) = \langle \Delta v(t) \Delta v(t + \tau) \rangle, 
    \quad \Delta v(t) =v(t) - \langle  v(t) \rangle,  
\end{equation}
can be obtained experimentally. Due to presence of the periodic driving of frequency $\omega$ this function is  periodic with respect to the first argument \cite{jung1993}
\begin{equation}
    \label{eq17}
    C(t, \tau) = C(t + \mathsf T, \tau)
\end{equation}
where $\mathsf T = 2\pi/\omega$ is a rescaled period of the ac current. Therefore we introduce the time average of the  autocorrelation function $C(t, \tau)$, namely
\begin{equation}
    \label{eq18}
    \mathcal{C}(\tau) = \frac{1}{\mathsf T} \int_0^{\mathsf T} dt \, C(t,\tau).
\end{equation}
The diffusion process is characterized by the low frequency part of the power spectrum of the voltage fluctuations \cite{atbook}
\begin{equation}
    \label{eq19}
    \chi(s) = \frac{1}{2} \int_{-\infty}^{+\infty} d\tau \, e^{is\tau} \mathcal{C}(\tau) \sim s^{1 - \alpha}.
\end{equation}
The diffusion coefficient is connected with the above equation via the relation \cite{atbook}
\begin{equation}
    \label{eq20}
    D_x = \chi(0) = \frac{1}{2} \int_{-\infty}^{+\infty} d\tau \, \mathcal{C}(\tau).
\end{equation}
Formulas (\ref{eq19}) and (\ref{eq20}) are particularly important from the experimental point of view as they allow for convenient measurement of both the nature of diffusion and if needed also its coefficient.

Finally, the ratio $D_x/2\pi$ can be considered as a velocity describing the normal phase diffusion over one period of the potential $U(x)$. Its relation to the average voltage $\langle v \rangle$ determines the dimensionless P\'{e}clet number defined as \cite{machura2005, machura2006, machura2010}
\begin{equation}
    \label{eq21}
    Pe = \frac{2\pi |\langle v \rangle|}{D_x}.
\end{equation}
A large P\'{e}clet number indicates a motion of mainly regular nature. If it is small then random or chaotic influences dominate the dynamics.



\section{Regime of globally maximal efficiency: Normal diffusion}
\label{results}
\begin{figure*}[t]
    \centering
    \includegraphics[width=0.33\linewidth]{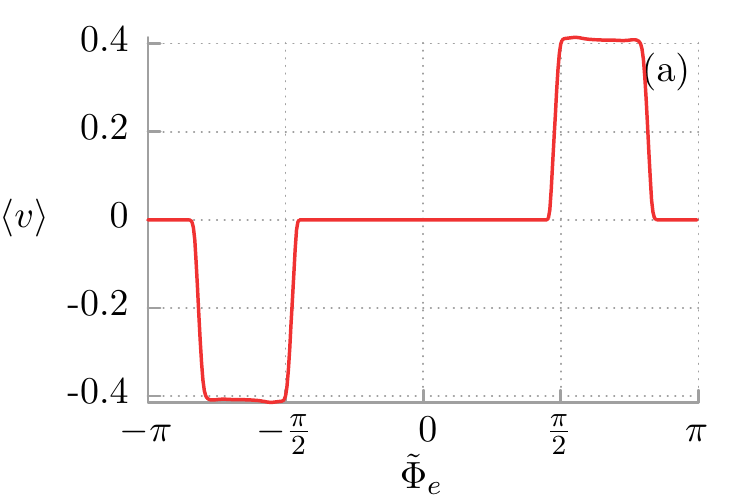}
    \includegraphics[width=0.33\linewidth]{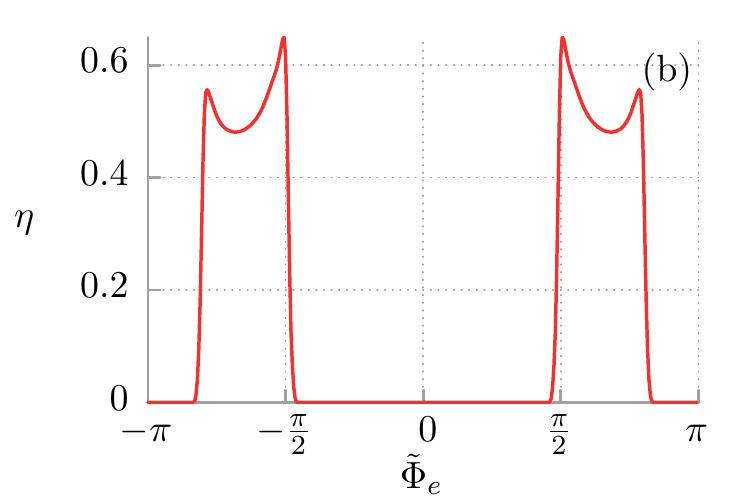} \\
    \includegraphics[width=0.33\linewidth]{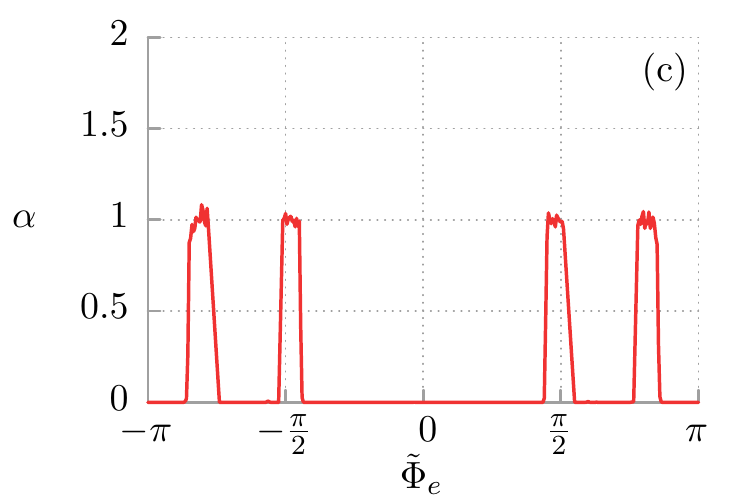}
    \includegraphics[width=0.33\linewidth]{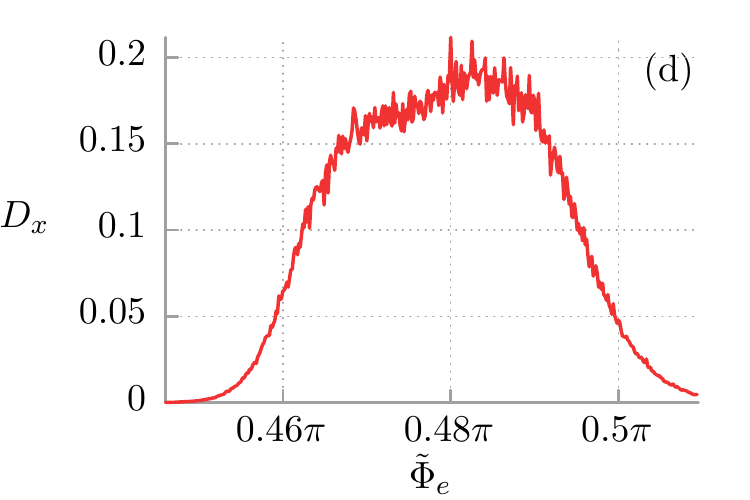}
    \caption{Impact of variation of the external magnetic flux $\tilde{\Phi}_e$ on relevant characteristics in the form of: (a) the directed transport measured as the dc voltage drop $\langle v \rangle$ across the SQUID, (b) the Stokes efficiency $\eta$, (c) phenomenological power exponent $\alpha$ describing phase diffusion in this setup and finally (d) the normal diffusion coefficient $D_x$. Other parameters are the same as in Fig. \ref{fig3}.}
    \label{fig5}
\end{figure*}


We have integrated the Langevin equation (\ref{eq6}) by employing a weak version of the stochastic second order predictor corrector algorithm \cite{platen} with a time step typically set to about $10^{-3} \cdot 2\pi/\omega$. Since Eq. (\ref{eq6}) is a second-order differential equation, we have to specify two initial conditions $x(0)$ and $\dot{x}(0)$.  We have chosen phases $x(0)$ and dimensionless voltages $\dot{x}(0)$ equally distributed over interval $[0,2\pi]$ and $[−2,2]$, respectively. All quantities of interest were ensemble-averaged over $10^3$ -- $10^4$ different trajectories which evolved over $10^3$ -- $10^4$ periods of the external ac driving. Numerical calculations were done by use of a CUDA environment implemented on a modern desktop GPU. This scheme allowed for a speed-up of a factor of the order $10^3$ times as compared to a common present-day CPU method.  \cite{januszewski2009, spiechowiczcpc} 


The system described by Eq. (\ref{eq6}) has a 7-dimensional parameter space $\{ \tilde{C}, a, \omega, F, j, \tilde{\Phi}_e, D \}$. Below, we study a nontrivial ratchet effect by putting the dc current $F = 0$.  We have performed scans of the parameter space at a high resolution to determine the general behaviour of the system. The conditions that are necessary for the generation and control of the direction of transport in the SQUID have been extensively studied in these regimes in our previous work.\cite{spiechowicz2014}  Moreover, very recently we also reported a regime for which the rectification efficiency $\eta$ is globally maximal. \cite{NJP15} In this regime the capacitance is $\tilde{C} 
\approx 0.496$, the amplitude of the ac current reads $a \approx 1.55$ and its is frequency $\omega \approx 0.406$. 
%
%
%
Let us now focus on the connection between the voltage drop $\langle v \rangle$, its rectification efficiency $\eta$ and the phase diffusion process across the SQUID in this prominent regime. 

In Fig. \ref{fig3} we present influence of the  ac driving amplitude $a$ on all characteristics describing quality of the transport process observed in the SQUID. In particular, panel (a) shows the dc voltage drop $\langle v \rangle$  across the device. There is a finite window of the parameter $a$ for which the directed transport of about equal phase velocity $\langle v \rangle$ is observed. When $a$ is smaller than $a \approx 1.53$ then the rocking mechanism is too weak to induce the non-negligible voltage drop. Similar situation can be discovered for the amplitudes greater than $a \approx 1.82$. These facts have their further consequences in the dependence of the Stokes efficiency $\eta$ on the ac driving amplitude $a$. Since this quantity is proportional to $\langle v \rangle^2$ it vanishes when there is no directed transport, see panel (b) of the same figure. The most intriguing feature of this plot is emergence of the rapid maximum for $a \approx 1.55$. It is caused by the evident local minimum of the voltage fluctuations $\sigma_v$ in the vicinity of this point. We refer the reader to Ref. \onlinecite{NJP15} for a detailed study on this effect. Panel (c) depicts the phenomenological power exponent $\alpha$ characterizing the diffusion process as a function of the ac driving amplitude $a$. It was computed from the slopes of time evolution of the mean squared displacement $\langle \Delta x^2 (t)\rangle$.  Surprisingly, apart from two clearly visible intervals of the ac driving amplitude where $\alpha \approx 1$ there is no phase diffusion. Especially, it is noteworthy that a wide window of $a$ can be observed where there is a finite dc voltage drop $\langle v \rangle \neq 0$ and simultaneously the power exponent vanishes $\alpha = 0$, yielding in a directed and non-diffusive transport across the device. In the neighbourhood of the  maximal efficiency for  $a \approx 1.55$   the exponent  $\alpha$ is guaranteed to be unity. In consequence, the phase diffusion is normal,  the diffusion coefficient $D_x$ has well established physical interpretation and can be conveniently computed by use of the formula (\ref{eq15}). Its variation is shown in the panel (d) of Fig. \ref{fig3}.  The most important finding is that transport in this  regime  is essentially stochastic. However,  the diffusion coefficient is small $D_x \approx 0.05$.   Moreover, it is seen that one can change the magnitude of $D_x$  by adjusting the ac current amplitude $a$. In particular, a small change of $a$  is accompanied by multiple increase  of  the diffusion coefficient $D_x$ thus leading to the phenomenon of diffusion   enhancement. \cite{constantini1999, reimann2001, reimann2002pre}
\begin{figure*}[t]
    \centering
    \includegraphics[width=0.3\linewidth]{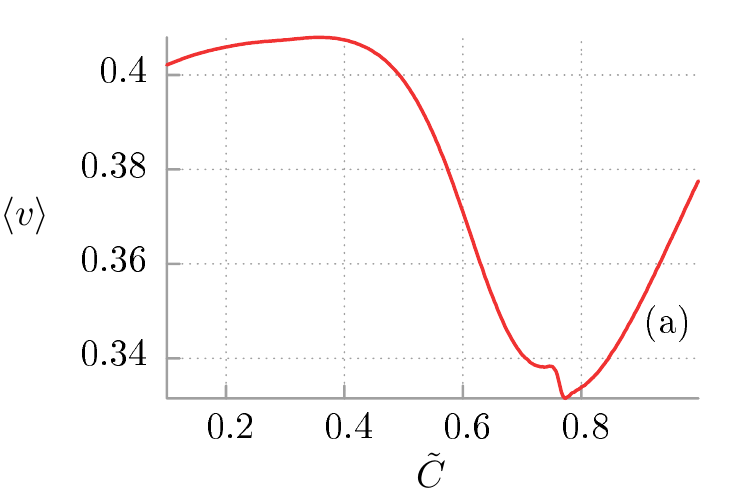}
    \includegraphics[width=0.3\linewidth]{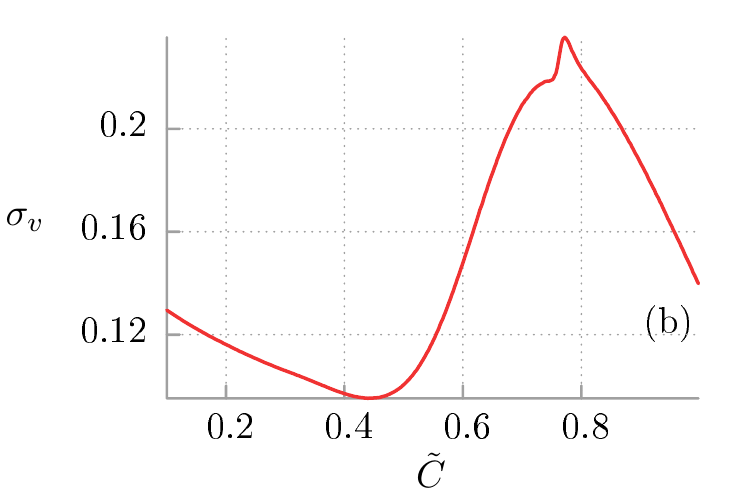} 
     \includegraphics[width=0.3\linewidth]{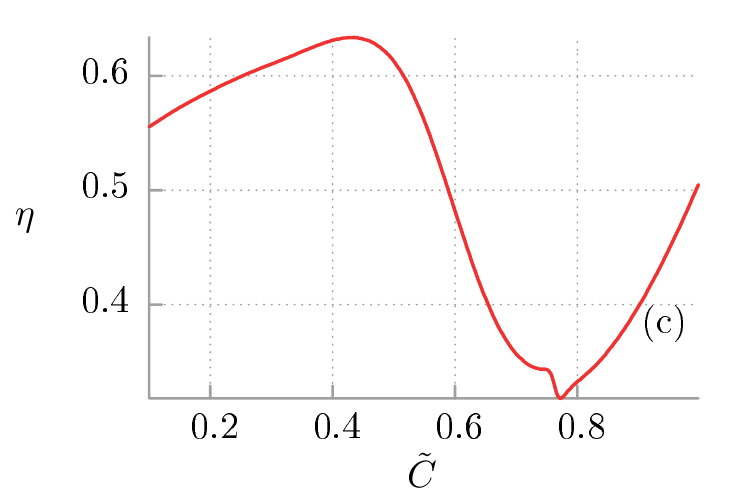} \\
    \includegraphics[width=0.3\linewidth]{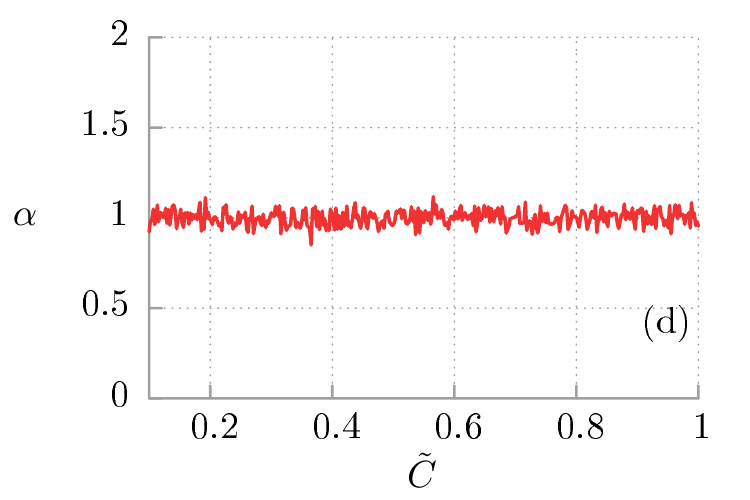}
    \includegraphics[width=0.3\linewidth]{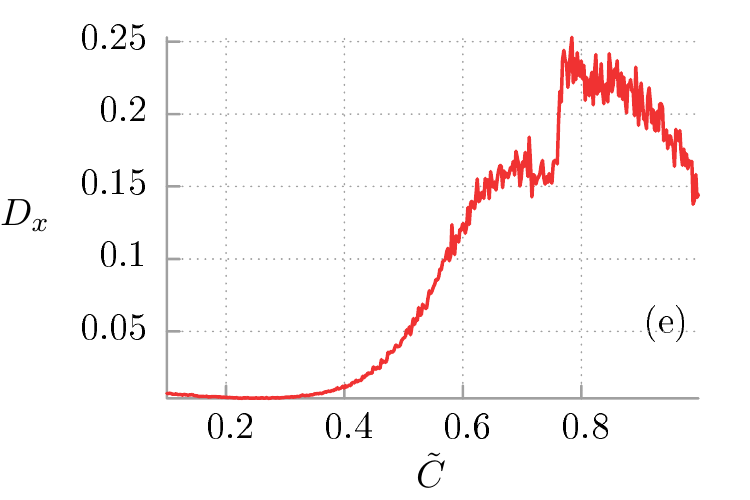}
    \caption{Influence of variation of the capacitance $\tilde{C}$ on relevant characteristics in the form of: (a) the directed transport measured as the dc voltage drop $\langle v \rangle$ across the SQUID, (b) the voltage fluctuations $\sigma_v$, (c) the Stokes efficiency $\eta$, (d) phenomenological power exponent $\alpha$ describing phase diffusion in this setup and finally (e) the normal diffusion coefficient $D_x$. Other parameters are the same as in Fig. \ref{fig3}.}
    \label{fig6}
\end{figure*}
\begin{figure}[b]
    \centering
    \includegraphics[width=0.7\linewidth]{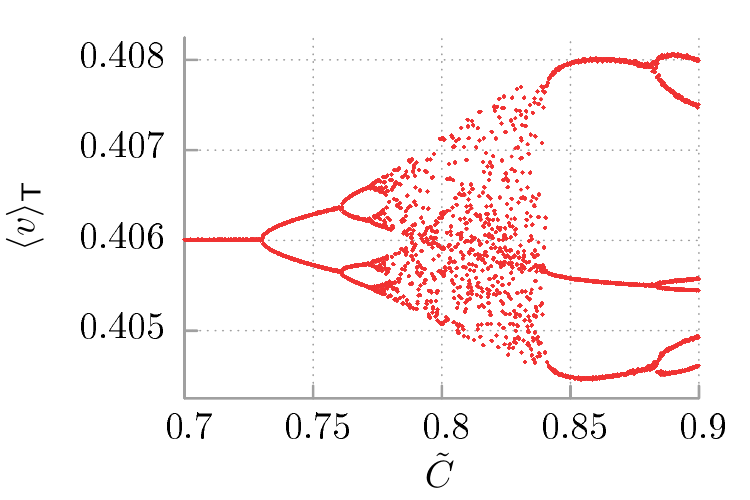}
    \caption{The dc voltage bifurcation diagram in the deterministic limit $D = 0$ as a function of the capacitance $\tilde{C}$ of the SQUID. The remaining parameters are given in Fig. \ref{fig3}.}
    \label{fig7B}
\end{figure}

To gain insight into the nature of the  phase diffusion process we have computed the corresponding deterministic ($D = 0$) dc voltage bifurcation diagram. In Fig. \ref{fig4} we show the asymptotic long time voltage $\langle v \rangle_{\mathsf T}$ averaged over the period $\mathsf T =2\pi/\omega$ of the external ac driving for 1024 different initial conditions $x(0)$ and $\dot{x}(0)$ randomly sampled from the intervals $[0,2\pi]$ and $[-2,2]$, respectively. Therefore all existing attractors are plotted at a given value of control parameter $a$. This figure reveals an unexpected simplicity of the phase space dynamics. The most important observation is that the system described by Eq. (\ref{eq6}) is non-chaotic in the analyzed regime  of the parameters. Moreover, there are regions for which only a single period one attractor exists, meaning that eventually all initial conditions evolve to it. Thus, each trajectory undergoes the same kind of motion resulting in the power exponent $\alpha =0$ (cf. Fig. \ref{fig3}c). Moreover, there are also intervals where several attractors coexist. This, in fact, is enough to observe the phenomenon of \emph{deterministic diffusion} \cite{blackburn1996, harish2002, tanimoto2002, atbook}. At sufficiently high temperature, the system will be typically ergodic with thermal fluctuations enabling stochastic escape events among coexisting deterministic separate attractors. In particular, transitions between neighbouring periodic solutions give rise to diffusive directed transport.

Let us now briefly discuss an influence of the external magnetic flux $\tilde{\Phi}_e$ on all previously introduced transport measures. From the symmetry considerations of Eq. (\ref{eq6}) it follows that for an arbitrary integer number $n$, the transformation $\tilde{\Phi}_e \rightarrow 2\pi n - \tilde{\Phi}_e$ reverses the sign of the average voltage $\langle v \rangle \rightarrow -\langle v \rangle$. This fact can be directly observed in panel (a) of Fig. \ref{fig5}. There are  two intervals where the average voltage drop assumes non-zero values which differ only with the direction of transport. Since the external magnetic flux alters the effective potential experienced by the  phase $x$  we conclude that in order to detect the average voltage one must tune it to a given rocking mechanism in the form of the ac driving of the amplitude $a$ and the frequency $\omega$. Similarly to the previously discussed case this fact is reflected in the dependence of the Stokes efficiency $\eta$ on the external magnetic flux $\tilde{\Phi}_e$. A careful inspection of panel (b) reveals that one can tune the effectiveness of voltage rectification just by correct adjustment of the external magnetic flux. The next plot depicts impact of  the external magnetic flux $\tilde{\Phi}_e$ on the  exponent $\alpha$.  Regions where the transport is non-diffusive dominate this parameter space. It is associated with existence of a single period one attractor describing either the running or locked solution of (\ref{eq6}) as it was in the previous case. Still, there are also intervals where the diffusive motion of the phase $x$ can be observed with $\alpha \approx 1$. In particular, panel (d) presents the dependence of the  diffusion coefficient $D_x$ on the external magnetic flux $\tilde{\Phi}_e$ for such a scenario: One can conveniently manipulate the phase diffusion by change of the external magnetic flux. This way of control of the diffusion process is very convenient from the experimental point of view. 
\begin{figure*}[t]
    \centering
    \includegraphics[width=0.33\linewidth]{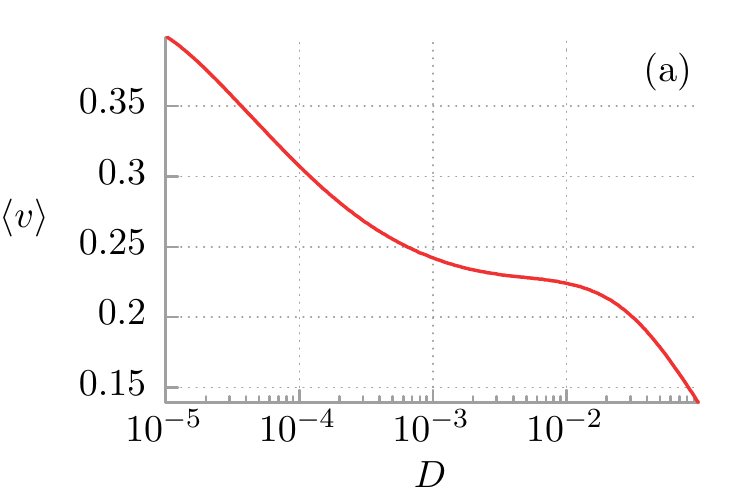}
    \includegraphics[width=0.33\linewidth]{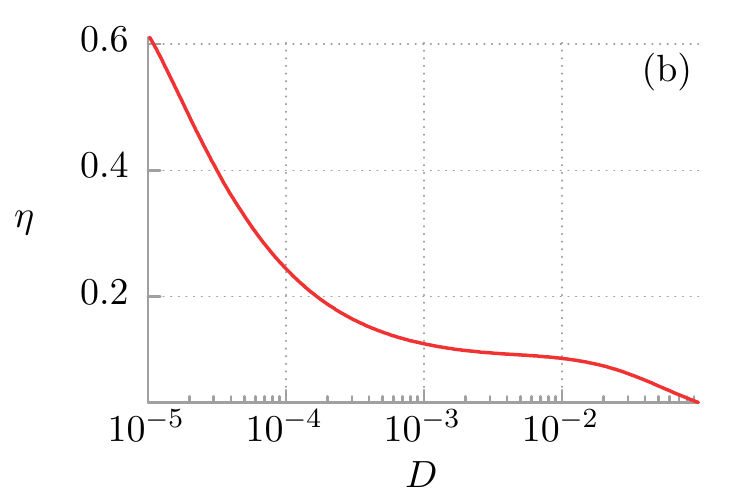} \\
    \includegraphics[width=0.33\linewidth]{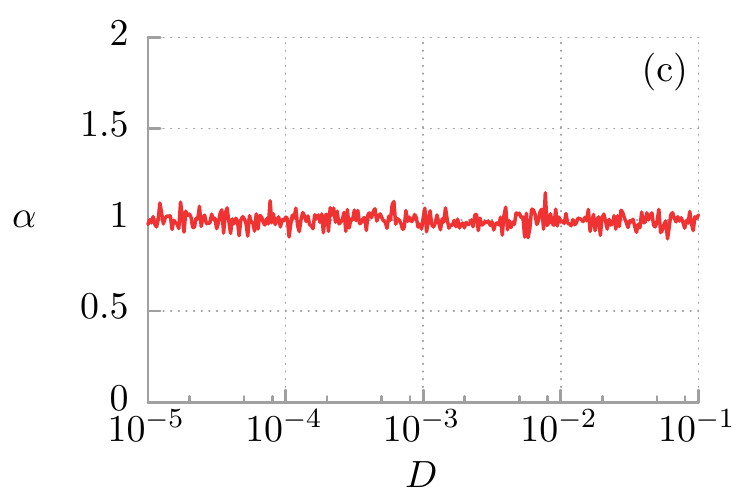}
    \includegraphics[width=0.33\linewidth]{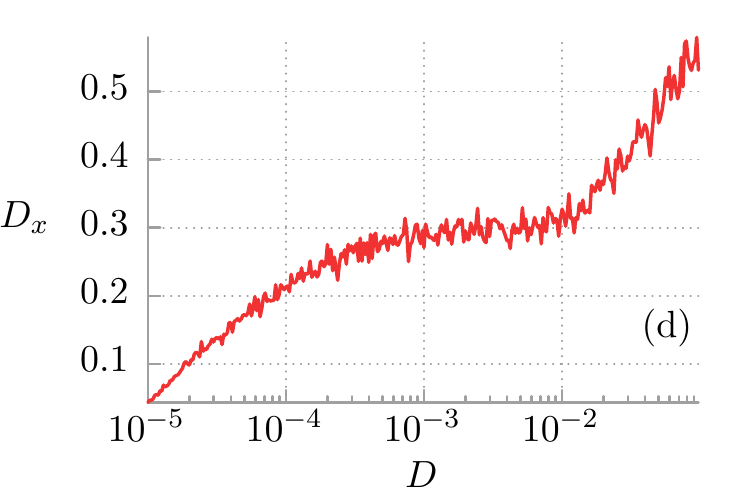}
    \caption{Panel (a) the directed transport measured as the dc voltage drop $\langle v \rangle$ across the SQUID, (b) the Stokes efficiency $\eta$, (c) phenomenological power exponent $\alpha$ describing phase diffusion in this setup and finally (d) the normal diffusion coefficient $D_x$ presented as a function of thermal noise intensity $D$. Other parameters are the same as in Fig. \ref{fig3}.}
    \label{fig7}
\end{figure*}

We now focus on the influence of inertia  described by  the capacitance $\tilde{C}$ of the SQUID on the phase diffusion, see  Fig. \ref{fig6}. Panels (a) and (c) depict the average voltage drop $\langle v \rangle$ across the device and the  Stokes efficiency $\eta$ as a function of the capacitance $\tilde{C}$. They are non-monotonic functions similar in shape but without any immediately obvious relation to each other. However, the most important observation is that in the vicinity of a point corresponding to the maximum transport efficiency $\eta$ also the voltage drop $\langle v \rangle$ is large. Notably, it lies closely to the border between overdamped $\tilde{C} \to 0$ and damped $\tilde{C} \approx 1$ regime. 
In panel (b), fluctuations of the phase velocity are depicted. One can notice an evident correlation, when velocity fluctuations are maximal, the efficiency is minimal. The next two panels present the diffusive behaviour as a function of the same parameter. In panel (d) one can see that the transport is essentially diffusive $\alpha \approx 1$ for entire interval of the inertial term variance. The last plot depicts the  diffusion constant $D_x$ versus the capacitance  $\tilde{C}$ of the device. In the overdamped limit the transport is rather  regular as the diffusion coefficient is very small. On the contrary, an increase of inertia is accompanied by simultaneous grow of the phase diffusion with a sharp increase in the vicinity $\tilde C \approx 0.78$. One can observe that when the average voltage $\langle v \rangle$ and Stokes efficiency $\eta$ are locally minimal and velocity fluctuations are maximal then the diffusion coefficient $D_x$ takes its maximum. Consequently, the transport is highly irregular and not optimal. Therefore we validate that quantities characterizing the phase diffusion in the device are somehow complementary to those usually used in order to describe the quality of transport. They give additional information which often corresponds well with the one measured by the voltage fluctuations (\ref{eq10}) or the efficiency (\ref{eq12}). It is intriguing to find a deeper reason of the sharp increase of $D_x$ in the vicinity $\tilde C \approx 0.78$. In Fig. \ref{fig7B}, we present the dc voltage  bifurcation diagram in the deterministic limit $D = 0$ as a function of the capacitance $\tilde C$ of the SQUID. In the vicinity of $\tilde C \approx 0.78$, a cascade of bifurcations is observed and a transition to chaos takes place. Due to this fact, the phase diffusion can be observed even in the purely deterministic regime and the diffusion coefficient rapidly increases.

Finally, we examine the influence of thermal fluctuations on the phase diffusion process. The relevant panels comparing all discussed quantities are presented in Fig. \ref{fig7}. One  can see  that an increase of the thermal fluctuation intensity $D$ leads to monotonic decrease of both the average voltage drop $\langle v \rangle$ across the SQUID and its voltage rectification efficiency $\eta$. Therefore, we conclude that it is a regime for which impact of thermal fluctuations  is destructive. It is confirmed in the last two panels of this figure where the phase diffusion process is studied. Not unexpectedly, the system is diffusive for the entire range of thermal fluctuation intensity $D$. It is because thermal noise activates stochastic transitions between coexisting deterministic disjoint attractors. Moreover, in the last panel we can see that when the temperature grows the transport becomes more and more diffusive as the diffusion coefficient $D_x$ monotonically increases.
\begin{figure*}[t]
    \centering
    \includegraphics[width=0.33\linewidth]{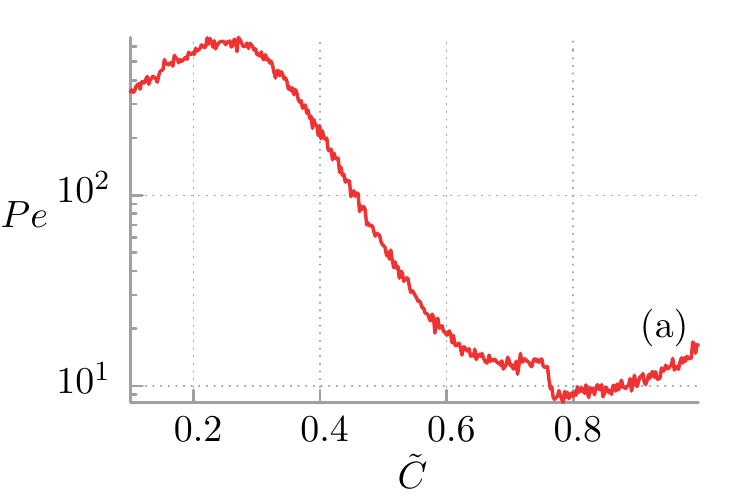}
    \includegraphics[width=0.33\linewidth]{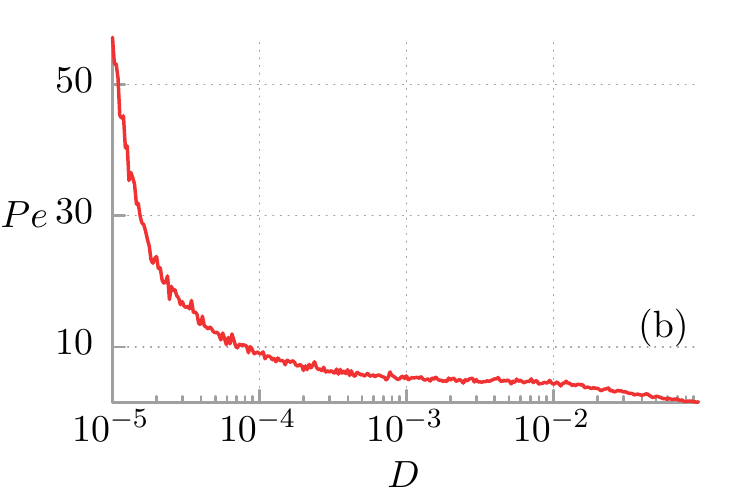}
    \caption{The dimensionless P\'{e}clet number is presented as a function of the capacitance $\tilde{C}$ of the SQUID and the thermal noise intensity $D$ in panel (a) and (b), respectively. Other parameters are listed in Fig. \ref{fig3}.}
    \label{fig8}
\end{figure*}

We collect part of our results by presenting the dependence of the dimensionless P\'{e}clet number on the SQUID capacitance $\tilde{C}$ and the thermal noise intensity $D$. The corresponding panels can be found in Fig. \ref{fig8}. In the first one the reader can observe clear separation between the overdamped $\tilde{C} \to 0$ and underdamped $\tilde{C} \approx 1$ regimes. In the former case the P\'{e}clet number is large indicating the transport of predominantly regular nature. Contrary, in the latter one it is small saying that the phase motion is chaotic or diffusive. Having in mind that this particular regime is essentially non-chaotic, we conclude that in that region of a parameter space  transport is diffusive. This fact support our previous statement on that subject. Moreover, the most important remark is that the P\'{e}clet number corresponding to the point $\tilde{C} = 0.496$ of globally maximal Stokes efficiency $\eta$ is relatively large being evidence of the diffusive but still highly regular transport. In panel (b) of the same figure we can observe a  fast monotonic decay of the P\'{e}clet number with increasing temperature of the  system. Therefore, in this regime thermal fluctuations have destructive impact on all  transport  quantifiers, starting from the average voltage $\langle v \rangle$, by the Stokes efficiency $\eta$ of the SQUID and finally up to irregularity of the phase motion reflected in the large diffusion coefficient $D_x$.
\section{Summary}
\label{summary}

We have studied diffusion process of the Josephson phase in the asymmetric SQUID system. Our analysis has been restricted to the regime of the maximal Stokes efficiency of the device and found normal diffusion with the exponent $\alpha \approx 1$. We have searched a neighbourhood of this point in the parameter space to check robustness of the observed behaviour in domains $\{a, \Phi_e, \tilde C, D \}$. When the amplitude $a$ of ac current is changed two windows of normal diffusion are detected (panel (c) in Fig. 4). In turn when the applied magnetic field is varied four intervals of normal phase motion are observed (panel (c) in Fig. 6). These areas are interrupted by windows where the exponent $\alpha =0$ and there is no diffusion. The dependence on the capacitance $\tilde C$ and intensity $D$ of thermal fluctuations is robust. The exponent $\alpha \approx 1$ for overdamped and damped case $\tilde C < 1$ (panel (d) in Fig. 7) and rescaled temperature $D$ which can change several orders of magnitude (panel (c) in Fig. 8). 

We have presented the possibility of convenient manipulation of the diffusion coefficient $D_x$  by tuning the experimentally accessible parameters of the setup like the ac driving amplitude $a$ or the external magnetic flux $\tilde{\Phi}_e$. Surprisingly, by doing it one can change its value several times. This lead us to the phenomenon of \emph{diffusion enhancement}.

Last but not least, we have found that the regime of maximal Stokes efficiency is essentially \emph{non-chaotic}. Regions where only a single period one attractor exists dominate the parameter space in the vicinity of this prominent area. Consequently, the phase motion is there non-diffusive. However, there are also intervals where due to the coexistence of several deterministic separate attractors the phase diffusion process can be observed. Then, sufficiently large thermal fluctuations enable stochastic transitions between them resulting in the diffusive directed transport. Overall, thermal noise has destructive impact on all presented quantities measuring the quality of transport in the system, starting from the average voltage $\langle v \rangle$ and ending on the diffusion coefficient $D_x$. It is remarkable that the non-diffusive or regular nature of the phase motion across the device is detected mainly in the overdamped regime $\tilde{C} \to 0$. Contrary, when the inertial term is increasing then the transport becomes more and more diffusive.

Finally, an interesting question concerns a possibility of observing the \emph{anomalous phase diffusion} \cite{atbook, metzler2000} in this setup. This question has been answered in the positive for the case of the extremely underdamped (hamiltonian) symmetric SQUID device in the deterministic limit. \cite{tanimoto2002} Our case is much more complicated, however, with the help of the computational power of modern GPU computers it should still be doable. This additional aspect is on our agenda for a potential future research.

The results described above may be helpful for further understanding of nontrivial response of nonlinear dynamics to external driving. They can readily be experimentally verified with an accessible setup consisting of three resistively and capacitively shunted Josephson junctions formed in an asymmetric SQUID device. Having in mind the mechanical interpretation of the studied model our research may also have potential applications for particle mixing, homogenization, selection or separation tasks. \cite{squires2005}

\section*{Acknowledgement}
This work was supported in part by the MNiSW program ”Diamond Grant” (J. S.) and NCN grant DEC-2013/09/B/ST3/01659 (J. {\L}.).

\end{document}